# USE OF GENOME INFORMATION-BASED POTENTIALS TO CHARACTERIZE HUMAN ADAPTATION

JAMES LINDESAY

*Department of Physics & Astronomy, Howard University, 2355 Sixth Street NW*
*Washington, DC 20059, United States*
*jlindesay@howard.edu*

TSHELA E MASON

*National Human Genome Center, Howard University, 2041 Georgia Avenue NW*
*Washington, DC 20060, United States*
*tshelamason@gmail.com*

WILLIAM HERCULES

*Department of Physics & Astronomy, Howard University, 2355 Sixth Street NW*
*Washington, DC 20059, United States*
*wmhercules@yahoo.com*

GEORGIA M DUNSTON

*Department of Microbiology, Howard University, 520 W Street NW*
*Washington, DC 20059, United States*
*gdunston@howard.edu*

As a living information and communications system, the genome encodes patterns in single nucleotide polymorphisms (SNPs) reflecting human adaption that optimizes population survival in differing environments. This paper mathematically models environmentally induced adaptive forces that quantify changes in the distribution of SNP frequencies between populations. We make direct connections between biophysical methods (e.g. minimizing genomic free energy) and concepts in population genetics. Our unbiased computer program scanned a large set of SNPs in the major histocompatibility complex region, and flagged an altitude dependency on a SNP associated with response to oxygen deprivation. The statistical power of our double-blind approach is demonstrated in the flagging of mathematical functional correlations of SNP information-based potentials in multiple populations with specific environmental parameters. Furthermore, our approach provides insights for new discoveries on the biology of common variants. This paper demonstrates the power of biophysical modeling of population diversity for better understanding genome-environment interactions in biological phenomenon.

---

**James Lindesay, Computational Physics Laboratory, Department of Physics & Astronomy, Howard University, 2355 Sixth Street NW, Washington, DC 20059, United States**

# 1. Introduction

As a complex, dynamic information system, the human genome encodes and perpetuates the principles of life. The information is incorporated within a mostly fixed template, as well as within the structure of human genome sequence variation. Of the approximately 3 billion nucleotides of the human genome, only about 0.1% consist of bi-allelic single nucleotide polymorphisms (SNPs) distributed throughout the genome[1]. Once the statistical distribution of variation reaches homeostasis in a given environment, a human *population* can be described in terms of the maintained order and patterns of polymorphisms in the whole genome. We define the environment not just in terms of geophysical parameters, but rather as the complete interface of the population to biologic and evolutionary influences. We assert that the stability of whole genome adaptation is reflected in the frequencies of maintained diversity in these common variants (SNPs) for a population in its environment.

As dynamic sites in the human genome, SNPs are often highly correlated into combinations referred to as *haploblocks* whose *haplotypes* are maintained throughout generations with fixed frequencies within a given population. Such combinations of SNPs are said to be in *linkage disequilibrium* (LD). This reflects that certain SNP allelic combinations never appear within the population, implying that only certain haplotypes are biologically viable and generationally maintained. In population dynamics, *viability* manifests as maintained survivability and functionality. *The formation of haploblocks is an emergent property of genomic information that cannot be characterized in the absence of the environmental influences that compel such phase transitions among populations.* Therefore, the dynamically independent statistical genomic units we use are SNP haplotypes together with alleles within SNP sites that are not in contiguous LD with any other SNPs.

In particular, changes in the distribution of allelic and haplotypic responses to the environment directly reflect adaptive forces on the population. The resilience of living humans as embodiments of the genome allows for the adaptation of groups to new or changing environments. Differing human populations have emerged as a consequence of the various past migratory groups remaining within specific environments, and developing the collective coping mechanisms that have allowed the groups to function effectively in their surroundings. We consider adaptation to be the dynamic process of modifying expressions of the genome towards optimizing the survivability of a group that remains in a particular environment. Using measures of genomic information that reflect the interplay of statistical variations due to the environmental baths within which stable populations exist motivates the development of 'genodynamics' as an analog to macro-physical 'thermodynamics'[2]. *This approach offers a novel way of thinking about population diversity, through the discovery of relationships between the environment and genome variation underlying biology.* In this paper, we mathematically model genome-environment interactions and demonstrate straightforward environmental influences upon genomic variants.

# 2. Materials and Methods

## 2.1. Population variation and information

We begin by developing expressions that relate the genomic information measures of human groups whose diversity profile is stable over generations, to additive dynamic state variables that depend upon the environment occupied by that group. Most common informatic measures in the physical and communication sciences are related to the entropy of the statistical system being described. In order to develop entropy measures for a genomic population, the dynamic units of relevance must first be ascertained. Within a given environment, the statistical distributions of certain sets of SNPs become highly correlated as emergent units. This means that the genomic information dynamics in a specific environment is an *emergent phase of expression* of the human genome.

The *specific entropy* $s^{(S)}$ (or per capita entropy) of a single SNP location *(S)* that is not in (contiguous) linkage disequilibrium will take the form of a canonical ensemble state variable in an environmental bath given by

$$s^{(S)} \equiv -\sum_{a=1}^{2} p_a^{(S)} \log_2 p_a^{(S)}, \qquad (1)$$

where $p_a^{(S)}$ represents the probability (frequency) that allele *a* occurs in the population. It should be noted that the entropy so defined is a dimensionless measure of disorder without biophysical units. Likewise, the specific entropy of a SNP haploblock *(H)* consisting of a set of strongly correlated bi-allelic SNPs is taken to be

$$s^{(H)} \equiv -\sum_{h=1}^{2^{n^{(H)}}} p_h^{(H)} \log_2 p_h^{(H)}, \quad (2)$$

where $n^{(H)}$ is the number of SNP locations in haploblock *(H)*, and $p_h^{(H)}$ represents the probability (frequency) that haplotype *h* occurs in the population. The upper limit in this sum represents the number of mathematically possible bi-allelic combinations of alleles within the haploblock. Commonly available tools were used to construct the haploblock structures[3].

Since entropy is a measure of the disorder of a distribution, a system with maximum disorder (equal statistical distribution of all mathematically possible combinations) is one of maximum entropy $S_{max}$. The *information content* (*IC*) of a maintained statistical distribution is measured by the degree of *order* that the distribution has relative to a completely disordered one, i.e., the difference between the entropy of a completely disordered distribution and that of the given distribution; $IC = S_{max} - S$[4]. Such an information measure is likewise additive due to the additive nature of the entropy[5]. Thus, both entropy and information content are extensive state variables whose values increase proportionate with the population size. The *normalized information content* (*NIC*) for a given SNP haploblock *(H)* is a (non-additive) intrinsic measure defined by

$$NIC^{(H)} \equiv \frac{S_{max}^{(H)} - S^{(H)}}{S_{max}^{(H)}} = \frac{s_{max}^{(H)} - s^{(H)}}{s_{max}^{(H)}} = \frac{n^{(H)} - s^{(H)}}{n^{(H)}}, \quad (3)$$

where, as previously stated the specific entropy of the haploblock $s^{(H)}$ is just the entropy per population member $S^{(H)} = N_{population} \, s^{(H)}$. This normalized measure of information ranges between 0 and 1. Such a dimensionless measure allows one to explore the informatic spectra of regions of the genomes of individuals as well as populations[4].

To best parameterize environmental influences, only Phase 3 HapMap data were utilized. These data include populations with African Ancestry in the Southwest USA (ASW), Utah residents with ancestry from Northern and Western Europe (CEU), Han Chinese in Beijing China (CHB), Chinese in Metropolitan Denver Colorado USA (CHD), Gujarati Indians in Houston Texas USA (GIH), Japanese in Tokyo Japan (JPT), Luhya in Webuye Kenya (LWK), Mexican Ancestry in Los Angeles California USA (MXL), Massai in Kinyawa Kenya (MKK), Toscani in Italia (TSI), and Yoruba in Ibadan Nigeria (YRI). Of the Phase 3 populations, the NIC of ASW is 0.52, CEU is 0.76, CHB is 0.76, GIH is 0.73, JPT is 0.77, LWK is 0.59, MXL is 0.71, MKK is 0.63, TSI is 0.74, and YRI is 0.63. It should be noted that for the so-called *founder* populations where the genotyping was more complete for Phase I, II, and III of HapMap data, the NIC values for CEU, CHB, and JPT are somewhat higher (0.88) than the NIC of 0.77 for YRI.

*2.2. Information dynamics of the human genome*

We next develop dimensional scales and units that can quantify the relative pliability and elasticity of information dynamics between various populations and regions of the genome of the same population, analogous to the additive energy units in the physical sciences. In contrast to the fundamental particles of micro-physics, fundamental life units cannot maintain in the absence of the environments that support them. Therefore, the least complicated description of genomic dynamics should develop genomic *free energy* variables $F_{Genome}$ as more fundamental than environmentally independent energetic measures.

The genomic free energy $F_{genome}$ has been developed as a state variable that balances between the conservation and variation of SNPs and haplotypes within a given environmental bath. Minimizing the genomic free energy optimizes the population's survivability under the complete set of environmental stimuli and stressors, establishing the balance between conservation and variation of alleles and traits in the dynamics of the population distribution. A dimensional environmental potential $T_E$ (which is an intensive state variable that is independent of the size of the population) will parameterize the intrinsic, pervasive agitation of the population due to stochastic environmental stimuli (analogous to how the *temperature* parameterizes agitation of fundamental physical units in a thermal bath).  Similarly, dimensional allelic and haplotype potentials, $\mu_a^{(S)}$ and $\mu_h^{(H)}$, will parameterize the genomic free energy change in a population from the addition of one individual of allele *a* or haplotype *h*.  For a given haploblock *(H)*, the differential genomic free energy takes the form

$$dF^{(H)} = -S^{(H)} dT_E + \sum_h \mu_h^{(H)} dN_h^{(H)} , \qquad (4)$$

where $N_h^{(H)}$ represents the number of individuals in the population with haplotype *h*.  This form neglects any influence of the population upon the environment.  The total genomic free energy is a sum over all SNP haploblocks and non-linked SNPs given by

$$F_{genome} = \sum_H F^{(H)} + \sum_S F^{(S)} . \qquad (5)$$

As is the case in thermodynamics, the additive allelic potentials $\mu_h^{(H)}$ are expected to scale relative to the environmental potential $T_E$, and allelic or haplotypic potential differences should directly reflect in the ratio of the frequencies of occurrence of those dynamic units within the population.  We assert that such properties are encompassed in the functional form

$$\frac{\mu_{h_2}^{(H)} - \mu_{h_1}^{(H)}}{T_E} = -\log_2 \frac{p_{h_2}^{(H)}}{p_{h_1}^{(H)}} . \qquad (6)$$

Defining a single human Genomic Energy Unit ($\widetilde{\mu} \equiv 1 \; GEU$) to be the allelic energy necessary to induce maximal variation within a single non-linked bi-allelic SNP location ($p_{a_1} = \frac{1}{2} = p_{a_2}$), the potential of the haplotype *h* or allele *a* in an environmental bath characterized by the environmental potential $T_E$ that bathes the whole genome can be expressed as

$$\begin{aligned} \mu_h^{(H)} &= (\widetilde{\mu} - T_E) n^{(H)} - T_E \log_2 p_h^{(H)} \\ \mu_a^{(S)} &= (\widetilde{\mu} - T_E) - T_E \log_2 p_a^{(S)} \end{aligned} . \qquad (7)$$

If only one allele is present at a SNP location for a given population, the allelic potential of that allele is defined to be at the *fixing potential* $\mu_{fixed}$ for that environment $\mu_{a_1}^{(S)} = \mu_{fixing} \equiv (\widetilde{\mu} - T_E)$.

We will assume that the population is homeostatic (or at least quasi-homeostatic, which means that any changes occurring in the population distribution requires many generations to become significant). Population homeostasis is equivalent to the Hardy-Weinberg condition used in population biology that the statistical distribution be independent of any sub-divisions of the population data, including those associated with differing generations or ages.  Our *population stability condition* will require that the genomic free energy be a (stable) minimum under changes in the population within the local environment when the population is in homeostasis with its environment, i.e., $\left( \frac{\partial F_{Genome}}{\partial N_{Population}} \right) = 0$.  By substituting the forms of the allelic potentials $\mu_h^{(H)}$ and $\mu_a^{(S)}$ expressed in terms of the probabilities in Eq. (7) into the population stability condition and summing over all haploblocks and SNPs, an explicit expression of the environmental potential can be obtained:

$$T_E = \frac{\widetilde{\mu} \, n_{SNPs}}{n_{SNPs} - s_{Genome}} = \frac{\widetilde{\mu}}{NIC_{Genome}} . \qquad (8)$$

This inversely relates the environmental potential to the intrinsic normalized information content characterizing the variation of the *whole* genome of the population, demonstrating that the whole genome is uniformly bathed in this particular environmental parameter.  The population stability condition can be expressed in terms of the population averaged haplotype and allelic potentials.  We refer to the average haplotype potential within a SNP haploblock $\sum_h \mu_h^{(H)} p_h^{(H)} = \langle \mu^{(H)} \rangle$ as the *block potential* for haploblock *(H)*, and the average allelic potential at a non-linked SNP location $\sum_a \mu_a^{(S)} p_a^{(S)} = \langle \mu^{(S)} \rangle$ as the *SNP potential* for location *(S)*. The population stability condition then requires that the sum of all block and SNP potentials for a given population vanishes:

$$\left(\frac{\partial F_{Genome}}{\partial N_{Population}}\right) = 0 \quad \Rightarrow \quad \sum_H \langle \mu^{(H)} \rangle + \sum_S \langle \mu^{(S)} \rangle = 0. \quad (9)$$

This condition demonstrates that balance is established between diversity and conservation in a population to optimize its survivability within the given environment. One should note that the environmental potential $T_E$, the block potentials $\langle \mu^{(H)} \rangle$ and the SNP potentials $\langle \mu^{(S)} \rangle$ can only be constructed for a population. In addition, the individual allelic potentials $\mu_h^{(H)}$ and $\mu_a^{(S)}$ characterize an overall allelic potential for each individual in the population,

$$\mu_{individual} = \sum_H \mu_h^{(H)} + \sum_S \mu_a^{(S)}, \quad (10)$$

where the set of SNP haplotypes *h* and alleles *a* are unique to the individual. An individual's overall allelic potential is not a universal parameter, but rather depends strongly upon the environment.

To illustrate population dependent spectra of genomic block potentials, the genomic free energies of blocks in the major histocompatibility complex (MHC) region on chromosome 6 are displayed for a few founder populations using phase I, II, and III data from HapMap in Figure 1.

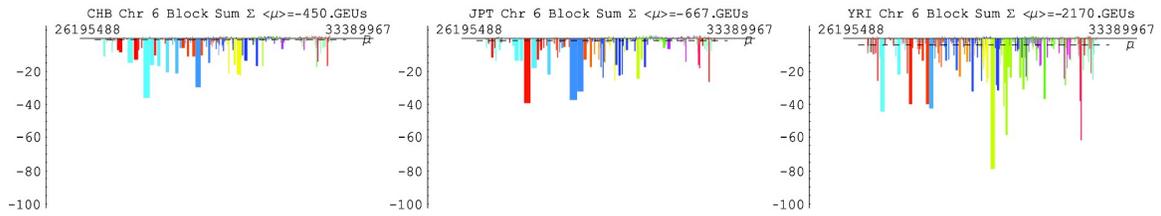

Figure 1: Block potentials for MHC region on Chromosome 6 as a function of location. Average values are demonstrated as the horizontal dashed lines. Regions of lower potential are indicative of a greater degree of conservation, and stronger binding of the correlated SNPs.

The MHC region encodes genes for the human immune response. This region of the genome is particularly relevant in host response to environmental stressors, and is known to display straightforward biological correlations with environmental parameters. The emergent differences in the haploblock structure of the populations are immediately apparent. The *block binding potential* (which parameterizes the stability of an emergent haploblock) will be defined as the difference in the block potential from the sum of the individual SNP potentials that make up that block if they were not in linkage disequilibrium (LD). The corresponding spectra of binding potentials (per SNP) are demonstrated in Figure 2.

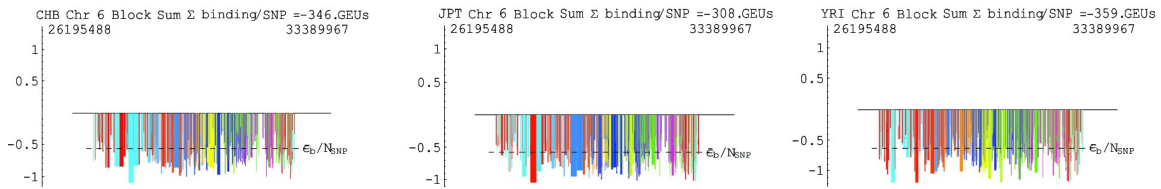

Figure 2: Binding block potential per SNP for the MHC region on Chromosome 6

Those SNPs in haploblocks with more negative binding potential per SNP have enhanced biologic favorability for maintaining their correlated statistics throughout generations of the populations in the given environments. SNPs in haploblocks with nearly zero binding potential per SNP are nearly independent, indicative of the environmental transition point of the emergent genomic phase. Stated precisely, an emergent genomic phase indicated by the formation of a haploblock of statistically correlated SNPs on the genome of a population in homeostasis with a particular environment results in a non-vanishing binding potential for the SNPs in that haploblock. *The strength of the binding block potential per SNP indicates the degree to which the SNP variation must be correlated in order to maintain a biologically viable population.*

*2.3. Distributive genodynamics*

The formulation of the information dynamics of the human genome in terms of genomic free energies directly results in well-defined forms for the SNP potentials for SNPs that are not in LD and for block potentials for correlated SNPs that are in LD. Since the SNP haploblock structure has an emergent form that differs between populations, meaningfully defined distributed potentials will reflect the biology underlying the participation of individual SNPs in the informatics architecture of its correlation with other SNPs in the haploblock. We will next develop *distributed SNP potentials* $\mu_S^{(H)}$ within a haploblock ($H$) such that they satisfy the following conditions:
- If the SNP is occupied by an allele that is fixed in the given population, then its distributed SNP potential is the fixing potential $\mu_{fixed}$;
- The sum of the distributed SNP potentials should be the same as the block potential $\mu^{(H)}$, i.e.

$$\langle \mu^{(H)} \rangle = \sum_{S=1}^{n^{(H)}} \mu_S^{(H)} \ ;$$

- The block potential should be linearly distributed amongst the constituent SNPs in accordance with occurrences of the SNP alleles.

The first bullet insures that if the SNP is not variant within the population, its genomic energy is not modified from that of a SNP that is not in LD, and the second bullet requires that the distributed potentials should reconstruct the block potential in an additive way. The third bullet represents a simple mechanism for relating the distributed potentials to the degree of variation in the SNP. The mathematical form that satisfies these conditions is given by

$$\mu_S^{(H)} \equiv \mu_{fixed} + \left[ \langle \mu^{(H)} \rangle - n^{(H)} \mu_{fixed} \right] \left( \frac{\overline{p}_S}{\sum_{S'} \overline{p}_{S'}} \right), \qquad (11)$$

where $\overline{p}_S = 1 - p_S$ is the minor allele frequency of the SNP labeled ($S$). Using this form, the distribution of the haploblock potential to any constituent SNP is proportionate to the occurrence of the minor allele in the population in a manner that increases the SNP's genomic free energy as the SNP has higher variation (i.e., becomes less conserved).

The degree of stability of the participation of the SNP in the biology of the emergent haploblock can be quantified in terms of its binding potential defined by

$$\varepsilon_{binding}^{(S)} \equiv \mu_S^{(H)} - \langle \mu^{(S)} \rangle, \qquad (12)$$

where $\langle \mu^{(S)} \rangle$ would be the SNP potential of the genomic variant were it not in LD. As defined, this metric of SNP binding within the haploblock is always negative, reflecting the increased genomic conservation inherent in LD.

We can furthermore assign allelic measures from the distributed SNP potentials in a manner that constructs the SNP potentials as population averages of derived *distributed allelic potentials* $\mu_{a_S}^{(H)}$, i.e., $\mu_S^{(H)} = \sum_{a_S} p_{a_S}^{(H)} \mu_{a_S}^{(H)}$. The most straightforward form that uniformly assigns the distributed SNP potential within a haploblock, and maintains the expected correlation that increased genomic potential reflects increased variation, results by simply adjusting the non-linked allelic potentials using the SNP binding potential, i.e.,

$$\mu_{a_S}^{(H)} \equiv \mu_{a_S}^{(S)} + \varepsilon_{binding}^{(S)} . \qquad (13)$$

It should be noted that all distributed potentials are only defined at the *population* level and cannot be ascribed to *individuals*. Only the emergent haplotype potentials $\mu_h^{(H)}$ can be ascribed to individuals within the population. However, since distributed potentials are defined for the population as a whole, they can be quite useful for parameterizing the environmental influences upon that population. *Distributed potentials are particularly useful for describing the adaptation of the population to stimuli and stressors with known biological correspondence to particular alleles or SNPs.* The description of genomic variants using distributed potentials inherently include any presently unknown *whole genome* response to specific stressors.

*2.4. Adaptive forces*

Once genomic free energy measures have been developed for individual alleles and genomic regions, environmentally induced adaptive forces can be characterized using gradients of those additive measures down the slope of environmental parameters. For a given allele *a* on the genome that is biologically connected to a definable environmental parameter $\lambda$ (such as UV light, lactose in diet, prevalence of malarial plasmodia, etc.), we define the environmentally induced *adaptive force* on that allele by

$$f_a \equiv -\frac{\partial \mu_a}{\partial \lambda}, \qquad (14)$$

with analogously defined adaptive forces on potentials characterizing SNPs, haploblocks, haplotypes, genes, and even perhaps whole chromosomes. Such an expression is only meaningful if there is a functional relationship between the biology of the genomic unit and the particular environmental parameter $\lambda$. In such cases, positive adaptive forces drive the conservation of the given genomic unit down the slope of the genomic potential. Increased *survivability* might drive the genomic unit towards more diversity, or more conservation, depending on the nature of the environmental influence upon the homeostatic population. Quantifying such forces inherently involves comparisons between differing environments.

To explore environmental impacts on adaptation, we will confine our investigation to phase III data of HapMap, since this represents the broadest set of populations with somewhat uniform genotyping. We have chosen to exclude ASW, CEU, CHD, GIH and MXL from our parameterization of adaptive forces, since these populations do not reside in their geographical origin. In this paper, the genomic potentials of the set of SNPs in the MHC region on chromosome 6 were chosen to conduct a double-blind exploration for possible correlations with three particularly straightforward environmental parameters: annual exposure to UV-B radiation, altitude above sea level, and exposure to malarial vectors. In order to simplify the analysis of any results, the set of all SNPs in this region that are *not* in LD for most of the populations were pre-selected out for the computational search. The algorithm examines whether the genomic potentials for the SNPs and alleles can be fitted to simple functional forms (curves) singly dependent on a given environmental parameter. If the root-mean-squared (RMS) deviation of the data points from the curves, as compared to the maximum variation of the data, falls within 10%, the SNP is flagged by the program, and adaptive forces are calculated for the curves.

The averaged ancestral annual UV-B radiation exposure used were expressed in units of Joules per square meter (UV radiance) as estimated from the following cited source[6]. In these units, estimates of annual UV radiance for the CHB population averaged 2180 (ranging from 1500 to 2600), for the JPT population averaged 2400 (ranging from 2300-2500), for the LWK population averaged 5764 (ranging from 5450 to 6500), for the MKK population averaged 5624 (ranging from 5000 to 6125), for the TSI population averaged 1507 (ranging from 950 to 2500), and for the YRI population averaged 5129 (ranging from 3500 to 6300). The altitude values used are averaged estimates of elevations of populated regions for ancestral homelands in units of meters using data from[7]. In units of meters, estimates of population elevation for the CHB population averaged 22 (ranging from 3 to 48), for the JPT population averaged 107 (ranging from 5 to 287), for the LWK population averaged 1711 (ranging from 1203 to 2486), for the MKK population averaged 1507 (ranging from 712 to 2383), for the TSI population averaged 74 (ranging from 1.3 to 143), and for the YRI population averaged 211 (ranging from 12 to 337). The parasite data was based upon the *Plasmodium falciparum* parasite rate (PfPR), used by the World Health Organization[8]. We expect that all of the examined populations had higher malarial exposure in ancestry than at present. In particular, the TSI population likely had significantly higher malarial exposure in ancestry than in present time, since relatively recent developments have significantly reduced the prevalence of the insects and treatment of the disease. In units of parasite reproductive rate, estimates of PfPR for the CHB population averaged 0.01 (ranging from 0 to 0.05), for the JPT populations averaged 0.0002 (ranging from 0 to 0.001), for the LWK population averaged 12 (ranging from 2 to 35), for the MKK population averaged 8 (ranging from 1 to 25), for the TSI population averaged 0.8 (ranging from 0 to 5), and for the YRI population averaged 70 (ranging from 20 to 95).

In the following plots, if there is a best fit curve plotted with the points, then the data was flagged by the computer program. Blue points represent populations with the flagged SNP not in linkage disequilibrium. The thickness of the curves in the plots represents the degree of correlation of the data with the fitting curve, with bolder curves indicating stronger correlations.

## 3. Results and Discussion

Our program flagged functional dependencies on altitude of phase III HapMap data for the SNP rs1109771 in the MHC region for the populations CHB, LWK, MKK, TSI and YRI. The curves are plotted in Figure 3.

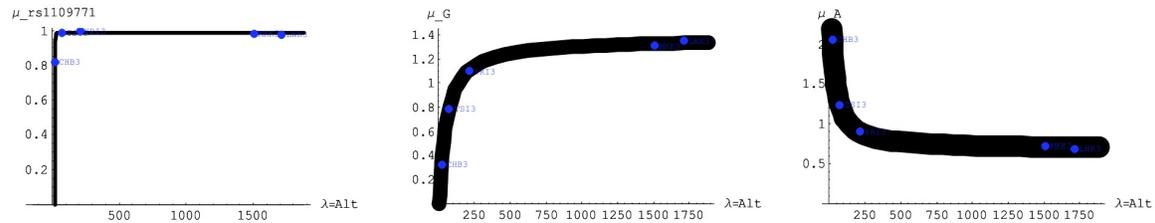
Figure 3: SNP rs1109771 in MHC region on Chromosome 6

The relative RMS deviation for the SNP potential was 0.03, for the $G$ allelic potential was 0.008, and for the $A$ allelic potential was 0.001. A significant adaptive force of about +1.5 GEUs/kilometer at lower altitudes on allele $A$ towards increased conservation is apparent. At higher altitudes, significant variation is maintained, as indicated by the SNP potential remaining very near the maximum value of 1 GEU (maximal variation). This implies that the $G$ allele continues a significant presence in the population in order to optimize its survivability in the higher altitudes available in the HapMap data.

### 3.1. High altitude and NOTCH4

Over the course of human history, adaptation to challenging environments has necessitated modulation of biological pathways at the genomic level to combat the toxic effects present in said environments. High altitude is an excellent example of how humans have adapted to an environmental stressor (e.g., low oxygen content). The body's response to chronic exposure to alveolar hypoxia is to hyperventilate, thereby increasing resting heart rate and stimulating the production of red blood cells to maintain the oxygen content of arterial blood at or above sea level values[9]. Moreover, an insufficient supply of oxygen prompts the formation of new vessels from the walls of existing ones, i.e. angiogenic sprouting[10]. Growth factors and chemokines are secreted from hypoxic tissues, stimulating endothelial cells to break away from vessel walls. These angiogenic factors then coordinate sprouting, branching, and new lumenized network formation until the oxygen content rises and normoxia can be re-established[11]. The Notch signaling pathway plays a key role in shaping the formation and remodeling of the vascular network under hypoxic conditions[10]. This pathway is an evolutionarily conserved intracellular signaling pathway that was originally identified in Drosophila. Notch has four transmembrane receptors, with Notch 1 and Notch 4 being expressed by endothelial cells[12-14]. It has been shown that targeted deletion of Notch 4 in mice results in the deregulation of arterial and venous specification of endothelial cells as well as the deformation of arteries and veins[15-16]. In addition, overexpression of the intracellular domain of Notch 4 in endothelial cells results in a β1 integrin-mediated increase in adhesion to collagen resulting in cells that show a reduced sprouting response to vascular endothelial growth factor both *in vitro* and *in vivo*[17]. Thus, it appears that Notch signaling promotes cellular responses in endothelial cells that help to alleviate the harmful effects of hypoxia in the human body. Consequently, population differences in allelic frequencies in this pathway could effectively provide an adaptive advantage for survival in response to this environmental stressor.

As a demonstration of the potential guidance offered by this formulation towards future discovery in the biology of whole genome adaptation, our program flagged functional dependencies on plasmodium parasite load from HapMap data for rs430620 in the MHC region for the populations CHB, LWK, MKK, TSI and YRI. The curves are plotted in Figure 4.

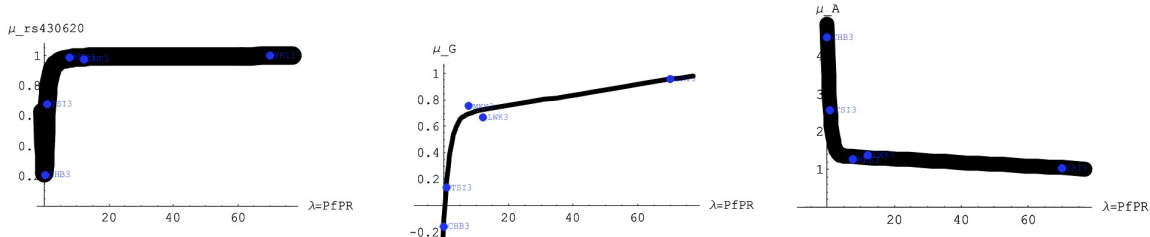

Figure 4: rs430620 in MHC region on Chromosome 6

This represents a strong flag for parasite dependency of a SNP in the intervening sequence of the genome with no known association to any gene. The relative RMS deviation for the SNP potential was 0.007, for the *G* allelic potential was 0.02, and for the *A* allelic potential was 0.008. A significant adaptive force of about +3 GEUs/unit PfPR for initial parasite loads on allele *A* towards increased conservation is apparent. The *A* allele has very low occurrence within populations with no parasite load, and the SNP approaches fixation towards allele *G*. Once again, for higher parasite loads, significant variation is maintained, as indicated by the SNP potential approaching the maximum of 1 GEU, indicative of the importance of maintaining a significant occurrence of the *G* allele in the population. The possibility of an association of the *A* allele with increased survivability under an environmental stressor that parallels this parasite load is intriguing.

## 4. Conclusion

We have demonstrated the utility of associating genomic free energy measures with environmental influences on whole genome adaptation. Double-blind smooth mathematical functions flagged relationships between altitude and the allelic energies of a SNP associated with oxygen deprivation. From these functional relationships, genomic energy gradients quantify adaptive forces in a manner analogous to corresponding concepts in the physical sciences. Our formulation of genomic information dynamics optimizes the survivability of a population in a given environment. Specifically, whole genome SNP distributions represent an environmentally influenced balance between genome sequence variation and conservation. Furthermore, double-blind smooth mathematical functions flagged relationships between parasite load and the allelic energies of a SNP with no known association to a gene. This provides an intriguing opportunity and direction for future discovery of the biology associated with this SNP.

## 5. Acknowledgements

The authors would like to acknowledge the continuing support of the National Human Genome Center, and the Computational Physics Laboratory, at Howard University. This research was supported in part by NIH Grant NCRR 2 G12 RR003048 from the RCMI Program, Division of Research Infrastructure. The authors hereby certify that they have no affiliations with or involvement in any organization or entity with any financial interest or non-financial interest in the subject matter or materials discussed in this manuscript.